\documentclass[apj]{emulateapj}
\usepackage{graphicx}
\usepackage{epsfig}
\usepackage{amssymb,amsmath}
\usepackage{array}
\usepackage{threeparttable}

\singlespace

\newcommand{\wmap}{{\sl WMAP}}
\newcommand{\planck}{{\it Planck}}
\newcommand{\herschel}{{\it Herschel}}

\newcommand{\ba}{\begin{eqnarray}}
\newcommand{\ea}{\end{eqnarray}}

\newcommand{\EV}[1]  {\langle#1\rangle}
\newcommand{\lmax}   {\ensuremath l_{\rm max}}

\newcommand  \beq    {\begin{equation}}

\newcommand  \eeq    {\end{equation}}
\newcommand  \gtsim  {\lower.5ex\hbox{$\; \buildrel > \over \sim \;$}}
\newcommand  \ltsim  {\lower.5ex\hbox{$\; \buildrel < \over \sim \;$}}

\newcommand{\LCDM}   {$\Lambda$CDM\,}

\newcommand{\lmaxval}   {3000}

\newcommand{\colorscaling}{1.11}

\newcommand{\AllpcBestfitTwothou}{\ensuremath  1.26^{+0.16}_{-0.20 }} 
\newcommand{\AllpcBestfitFourthou}{\ensuremath 0.94  \pm 0.08  }

\newcommand{\chicmb}{\chi_{\rm CMB}}

\usepackage{color}
\definecolor{orange}{rgb}{1,0.3,0}

\shortauthors{}
\shorttitle{ACTPol Lensing-CIB Cross-correlation}
\slugcomment{Draft: \today}

\begin{document}

\newcommand{\AllpcSig}{ 9.1} 
\newcommand{\AllpcBestfit}{\ensuremath 1.02^{+0.12}_{-0.08 }} 
\newcommand{\AllpcChisquare}{ 37.2}  
\newcommand{\AllpcPte}{ 0.14}  

\newcommand{\PolonlySig}{ 4.5} 
\newcommand{\PolonlyBestfit}{\ensuremath 1.26^{+0.28}_{-0.24 }} 
\newcommand{\PolonlyChisquare}{ 30.4}  
\newcommand{\PolonlyPte}{ 0.39}  

\newcommand{\EBSig}{ 3.2} 
\newcommand{\EBBestfit}{\ensuremath 1.30  \pm 0.40  } 
\newcommand{\EBChisquare}{ 25.9}  
\newcommand{\EBPte}{ 0.63}  

\newcommand{\dofsAllPatches}{29}

\title{The Atacama Cosmology Telescope: Lensing of CMB Temperature and Polarization Derived from Cosmic Infrared Background Cross-Correlation}

\author{
Alexander~van~Engelen\altaffilmark{1,2},
Blake~D.~Sherwin\altaffilmark{3},
Neelima~Sehgal\altaffilmark{2},
Graeme~E.~Addison\altaffilmark{4},
Rupert~Allison\altaffilmark{5},
Nick~Battaglia\altaffilmark{6},
Francesco~de~Bernardis\altaffilmark{7},
J.~Richard~Bond\altaffilmark{1},
Erminia~Calabrese\altaffilmark{5},
Kevin~Coughlin\altaffilmark{8},
Devin~Crichton\altaffilmark{9},
Rahul~Datta\altaffilmark{8},
Mark~J.~Devlin\altaffilmark{10},
Joanna~Dunkley\altaffilmark{5},
Rolando~D\"{u}nner\altaffilmark{11},
Emily~Grace\altaffilmark{12},
Megan~Gralla\altaffilmark{9},
Amir~Hajian\altaffilmark{1},
Matthew~Hasselfield\altaffilmark{13,4},
Shawn~Henderson\altaffilmark{7},
J.~Colin~Hill\altaffilmark{14},
Matt~Hilton\altaffilmark{15},
Adam~D.~Hincks\altaffilmark{4},
Ren\'ee~Hlozek\altaffilmark{13},
Kevin~M.~Huffenberger\altaffilmark{16},
John~P.~Hughes\altaffilmark{17},
Brian~Koopman\altaffilmark{7},
Arthur~Kosowsky\altaffilmark{18},
Thibaut~Louis\altaffilmark{5},
Marius~Lungu\altaffilmark{10},
Mathew~Madhavacheril\altaffilmark{2},
Lo\"ic~Maurin\altaffilmark{11},
Jeff~McMahon\altaffilmark{8},
Kavilan~Moodley\altaffilmark{15},
Charles~Munson\altaffilmark{8},
Sigurd~Naess\altaffilmark{5},
Federico~Nati\altaffilmark{19},
Laura~Newburgh\altaffilmark{20},
Michael~D.~Niemack\altaffilmark{7},
Michael~R.~Nolta\altaffilmark{1},
Lyman~A.~Page\altaffilmark{12},
Christine~Pappas\altaffilmark{12},
Bruce~Partridge\altaffilmark{21},
Benjamin~L.~Schmitt\altaffilmark{10},
Jonathan~L.~Sievers\altaffilmark{22,23,12},
Sara~Simon\altaffilmark{12},
David~N.~Spergel\altaffilmark{13},
Suzanne~T.~Staggs\altaffilmark{12},
Eric~R.~Switzer\altaffilmark{24,1},
Jonathan~T.~Ward\altaffilmark{10},
Edward~J.~Wollack\altaffilmark{24}
}
\altaffiltext{1}{Canadian Institute for Theoretical Astrophysics, University of
Toronto, Toronto, ON, Canada M5S 3H8}
\altaffiltext{2}{Physics and Astronomy Department, Stony Brook University, Stony Brook, NY USA 11794}
\altaffiltext{3}{Berkeley Center for Cosmological Physics, LBL and
Department of Physics, University of California, Berkeley, CA, USA 94720}
\altaffiltext{4}{Department of Physics and Astronomy, University of
British Columbia, Vancouver, BC, Canada V6T 1Z4}
\altaffiltext{5}{Sub-Department of Astrophysics, University of Oxford, Keble Road, Oxford, UK OX1 3RH}
\altaffiltext{6}{McWilliams Center for Cosmology, Carnegie Mellon University, Department of Physics, 5000 Forbes Ave., Pittsburgh PA, USA, 15213}
\altaffiltext{7}{Department of Physics, Cornell University, Ithaca, NY, USA 14853}
\altaffiltext{8}{Department of Physics, University of Michigan, Ann Arbor, USA 48103}
\altaffiltext{9}{Dept. of Physics and Astronomy, The Johns Hopkins University, 3400 N. Charles St., Baltimore, MD, USA 21218-2686}
\altaffiltext{10}{Department of Physics and Astronomy, University of
Pennsylvania, 209 South 33rd Street, Philadelphia, PA, USA 19104}
\altaffiltext{11}{Departamento de Astronom{\'{i}}a y Astrof{\'{i}}sica, Pontific\'{i}a Universidad Cat\'{o}lica,
Casilla 306, Santiago 22, Chile}
\altaffiltext{12}{Joseph Henry Laboratories of Physics, Jadwin Hall,
Princeton University, Princeton, NJ, USA 08544}
\altaffiltext{13}{Department of Astrophysical Sciences, Peyton Hall, 
Princeton University, Princeton, NJ USA 08544}
\altaffiltext{14}{Dept. of Astronomy, Pupin Hall, Columbia University, New York, NY, USA 10027}
\altaffiltext{15}{Astrophysics and Cosmology Research Unit, School of Mathematics, Statistics and Computer Science, University of KwaZulu-Natal, Durban 4041, South Africa}
\altaffiltext{16}{Department of Physics, Florida State University, Tallahassee FL, USA 32306}
\altaffiltext{17}{Department of Physics and Astronomy, Rutgers, 
The State University of New Jersey, Piscataway, NJ USA 08854-8019}
\altaffiltext{18}{Department of Physics and Astronomy, University of Pittsburgh, 
Pittsburgh, PA, USA 15260}
\altaffiltext{19}{Dipartimento di Fisica, Universit\`{a} La Sapienza, P. le A. Moro 2, 00185 Roma, Italy}
\altaffiltext{20}{Dunlap Institute for Astronomy and Astrophysics, University of Toronto, 50 St. George St., Toronto ON, Canada M5S 3H4}
\altaffiltext{21}{Department of Physics and Astronomy, Haverford College,
Haverford, PA, USA 19041}
\altaffiltext{22}{Astrophysics and Cosmology Research Unit, School of Chemistry and Physics, University of KwaZulu-Natal, Durban 4041, South Africa}
\altaffiltext{23}{National Institute for Theoretical Physics (NITheP), University of KwaZulu-Natal, Private Bag X54001, Durban 4000, South Africa}
\altaffiltext{24}{NASA/Goddard Space Flight Center, Greenbelt, MD, USA 20771}
\altaffiltext{25}{Department of High Energy Physics, Argonne National Laboratory, 9700 S Cass Ave, Lemont, IL USA 60439}
\altaffiltext{26}{Department of Astronomy, University of Illinois at Urbana-Champaign, W. Green Street, Urbana, IL, USA, 61801}
\altaffiltext{27}{School of Earth and Space Exploration, Arizona State University, Tempe, AZ, USA 85287}
\altaffiltext{28}{Department of Physics , West Chester University 
of Pennsylvania, West Chester, PA, USA 19383}
\altaffiltext{29}{NIST Quantum Devices Group, 325
Broadway Mailcode 817.03, Boulder, CO, USA 80305}
\altaffiltext{30}{Department of Physics, Stanford University, Stanford, CA, 
USA 94305-4085}
\altaffiltext{31}{Sociedad Radiosky Asesor\'{i}as de Ingenier\'{i}a Limitada Lincoy\'{a}n 54,
Depto 805 Concepci\'{o}n, Chile}
\altaffiltext{32}{School of Physics and Astronomy, Cardiff University, The Parade, 
Cardiff, Wales, UK CF24 3AA}
\altaffiltext{33}{National Center for Supercomputing Applications (NCSA), University of Illinois at Urbana-Champaign, 1205 W. Clark St., Urbana, IL, USA, 61801}


\begin{abstract}
We present a measurement of the gravitational lensing of the Cosmic Microwave Background (CMB) temperature and polarization fields obtained by cross-correlating the reconstructed convergence signal from the first season of ACTPol data at 146~GHz with Cosmic Infrared Background (CIB) fluctuations measured using the \planck\ satellite.  Using an overlap area of 206 square degrees, we detect gravitational lensing of the CMB polarization by large-scale structure at a statistical significance of $\PolonlySig \sigma$. Combining both CMB temperature and polarization data gives a lensing detection at $\AllpcSig \sigma$ significance.  A B-mode polarization lensing signal is present with a significance of $\EBSig \sigma$.  We also present the first measurement of CMB lensing--CIB correlation at small scales corresponding to $ l > 2000$.  Null tests and systematic checks show that our results are not significantly biased by astrophysical or instrumental systematic effects, including Galactic dust.  Fitting our measurements to the best-fit lensing-CIB cross power spectrum measured in \planck\ data, scaled by an amplitude $A$, gives $A=\AllpcBestfit$(stat.)$\pm 0.06$(syst.), consistent with the \planck\ results.
\end{abstract}


\section{Introduction}
\label{sec:intro}
\setcounter{footnote}{0} 

The Cosmic Microwave Background (CMB) temperature power spectrum has provided a wealth of information about the composition and evolution of the Universe \citep[e.g.,][]{spergel/etal/2003, story/etal/2013, hinshaw/etal/2013, das/etal/2014, planck_params/2013}.  More recently, measurements of the CMB polarization power spectrum have offered additional cosmological information (e.g., in the last five years: \citealp{brown/etal/2009, quiet-w/2012, barkats/etal/2014} [BICEP1 Collaboration],  \citealp{bicep2a/2014}, \citealp{naess/etal/2014} [ACTPol collaboration], \citealp{crites/etal/2014} [SPTpol collaboration]).  Gravitational lensing of the CMB has also emerged as another powerful means to place constraints on the late-time relationship between luminous and dark matter and to understand the evolution of large-scale structure \citep[e.g.,][]{bernardeau/1997, seljak/1999}.  Measurements of the lensing of the polarized CMB have become feasible only recently \citep{hanson/etal/2013, pbear-herschel/2013, pbear-eeeb/2013}, and with that has emerged a clean and powerful probe of neutrino properties and dark energy.  In this work, we present a detection of lensing of the CMB polarization field using data from the Atacama Cosmology Telescope Polarimeter (ACTPol) in cross-correlation with the Cosmic Infrared Background (CIB) measured using {\it Planck}/HFI (\planck\ hereafter). 

Lensing of the CMB temperature field has rapidly progressed from first detections \citep{smith/etal/2007, hirata/etal/2008, das/etal/2011, vanengelen/etal/2012} to precision measurements \citep{planck_lensing/2013}.  While lensing measures from the CMB alone provide direct constraints on the evolution of gravitational potentials, 
cross-correlations with tracers of large-scale structure have the advantage of being less sensitive to systematic errors and have potentially larger detection significance.  Indeed, the first detections of CMB lensing were obtained through cross-correlation with radio  and optical galaxies \citep{smith/etal/2007, hirata/etal/2008}.  These have been followed more recently by  cross-correlations with infrared-selected galaxies \citep{bleem/etal/2012, planck_lensing/2013}, submillimeter-selected galaxies \citep{bianchini/etal/2014},  quasars \citep{sherwin/etal/2012, planck_lensing/2013, geach/etal/2013, dipompeo/etal/2014}, gamma-rays \citep{fornego/etal/2014}, the cosmic shear from optical galaxies \citep{hand/etal/2013}, galaxy clusters emitting in the X-ray \citep{planck_lensing/2013} and via the Sunyaev-Zel'dovich effect \citep{hill/spergel/2013}, and the emission from unresolved dusty galaxies comprising the CIB \citep{holder/etal/2013, planck_ciblensing/2013, hanson/etal/2013, pbear-herschel/2013}.
The CIB at submillimeter and millimeter wavelengths has a flux distribution that peaks around a redshift of $z \sim 1.5$--$2$ \citep[e.g.,][]{viero/etal/2013, bethermin/etal/2013, addison/etal/2012}, and has substantial overlap with the redshift distribution of CMB lensing.  Recent measurements suggest that the correlation between CMB lensing and the CIB at 545~GHz is as large as $ 80\%$ \citep{planck_ciblensing/2013}.

Lensing of the polarized CMB can yield measurements of the projected matter density beyond the precision possible with lensing of the CMB temperature field.  This is because measurements of lensing-induced B-mode polarization are not limited by cosmic variance from primordial temperature perturbations \citep{hu/okamoto/2002}.  In addition, some sources of bias for CMB temperature lensing such as  galaxies and galaxy clusters \citep{osborne/etal/2013, vanengelen/etal/2014} are expected to have smaller relative signals in polarization maps \citep{smith/etal/2009}.  Detections of polarization lensing have been reported by the SPTpol \citep{hanson/etal/2013}, \textsc{Polarbear} \citep{pbear-eeeb/2013, pbear-herschel/2013}, and BICEP2 \citep{bicep2a/2014} teams.  This work presents the first measurements of polarization lensing using ACTPol, in cross-correlation with maps of the CIB from {\it Planck}.  These polarization lensing measurements are obtained from a wider composite sky area (206 deg$^2$) than previous measurements.
This paper also presents both CMB temperature and polarization lensing measurements at smaller angular scales than previously reported, allowing for novel tests of both CIB models and standard \LCDM cosmology.  

In this paper we use a fiducial cosmological model based on a fit to WMAP, SPT and ACT data \citep{calabrese/etal/2013}.  

\section{Data}
\label{sec:data}

\subsection{CMB Data}
\label{sec:cmb_data}

ACT is located at an altitude of 5190 m in Parque Astron\'omico Atacama in northern Chile.  The 6-meter primary mirror provides arcminute resolution at millimeter wavelengths.  Its polarization-sensitive camera, ACTPol, is described in \citet{niemack/etal/2010} and \citet[][N14 hereafter]{naess/etal/2014}.  ACTPol observed at 146~GHz from Sept.~11 to Dec.~14, 2013.   
Observations focused on four fields near the celestial equator at right ascensions of $150^\circ$, $175^\circ$, $355^\circ$, and $35^\circ$, which we call D1 (73 deg$^2$), D2 (70 deg$^2$), D5 (70 deg$^2$), and D6 (63 deg$^2$).  The scan strategy allows for each patch to be observed in a range of different parallactic angles while the telescope scans horizontally.  This aids in separating instrumental from celestial polarization.  The white noise map sensitivities in temperature for the patches are 16.2, 17, 13.2, and 11.2 $\mu$K-arcmin respectively, with polarization noise levels roughly $\sqrt{2}$ larger.  The beam size, measured as the full-width at half-maximum, is approximately 1.4$\arcmin$.     The patches were observed in sequence throughout day and night, with the nighttime data fractions  being $50\%, 25\%, 76\%$, and $94\%$ for D1, D2, D5, and D6 respectively.  In this analysis we use only nighttime data from D1, D5, and D6, which amounts to roughly 600 hours of observations.  The maximum-likelihood maps are made with 30 arcsecond
resolution.  Further details about the observations and mapmaking can be found in N14.

To treat point sources, we filter the D1, D5, and D6 patches with an optimal filter matched to the ACTPol beam profile, and identify point sources in the temperature with a signal five times larger than the mean background uncertainty  in the filtered maps.  By measuring the flux of each source, a template of beam-convolved point sources is constructed for each patch, which is then subtracted from the corresponding ACTPol patch.  In this way, point sources with fluxes above 8~mJy are removed from D1, and sources with fluxes above 5~mJy are removed from D5 and D6.  We also identify galaxy clusters by using a matched filter technique; for each cluster identified at signal-to-noise ratio greater than 4 we interpolate with a disk of radius $5\arcmin$ using the inpainting technique described by \citet{bucher/louis/2012}.

The ACTPol patches are calibrated to the \planck\ 143~GHz temperature map \citep{planck_mission/2013}  following the method described in \citet{louis/etal/2014}.  However, the patches are then multiplied by a factor of 1.012 to correspond to the \wmap\ calibration as in N14.

We construct an apodization window for each patch by tapering, with a cosine taper of width of 100\arcmin, the corresponding smoothed inverse variance weight map of that patch.  We multiply the T, Q, and U maps of each patch by this apodization window, effectively downweighting the noisier regions of the map. We then transform the Q and U maps into E and B maps following the pure-EB transform \citep{smith/2006, smith/zaldarriaga/2007, louis/etal/2013}.   The weighting and apodization reduces the effective area used in the analysis by 23\% compared to the area of the unwindowed patches.

\subsection{CIB Data}
\label{sec:cib_data}
For CIB maps, we use data obtained at 545~GHz using the \planck\ satellite that overlap the ACTPol D1, D5, and D6 survey regions.  The \planck\ maps for the nominal mission were retrieved from the \planck\ Legacy Archive \citep{planck_mission/2013}.  We clean these maps of bright point sources by masking extragalactic sources identified in the \planck\ point-source catalog as being above $5\sigma$ at 545~GHz.  Dusty Galactic emission is also cleaned from the \planck\ maps by masking the maps based on HI maps obtained from the  Leiden/Argonne/Bonn survey \citep{land/slosar/2007}, removing regions with HI column density greater than $3.6\times10^{20}$cm$^{-2}$.  This corresponds to  36, 30, and 51 square degrees, or fractions of  {46\%, 53\%, and 73\% of  D1, D5, and D6, respectively.  We also deconvolve the \planck\ maps with a beam which we take to be a Gaussian profile with width $\theta_{\rm FWHM} = 4.84'$ \citep{planck_beams/2013}, and apply the same apodization window that we apply to the ACTPol data.

\section{Lensing and Cross-Correlation Pipeline}
\label{sec:pipeline}

Gravitational lensing by large-scale structure deflects the paths of CMB photons by the angles given by the gradient of the projected gravitational potential (i.e., ${\bf d} = \nabla \phi$ where ${\bf d}$ is the deflection field and $\phi$ is the projected potential).  This re-mapping of the primordial CMB correlates previously independent pairs of modes, and also converts E-modes into B-modes.  This creates correlations between E and B modes which would otherwise be independent.  
\begin{figure}[t]
\includegraphics[width=1.05\columnwidth]{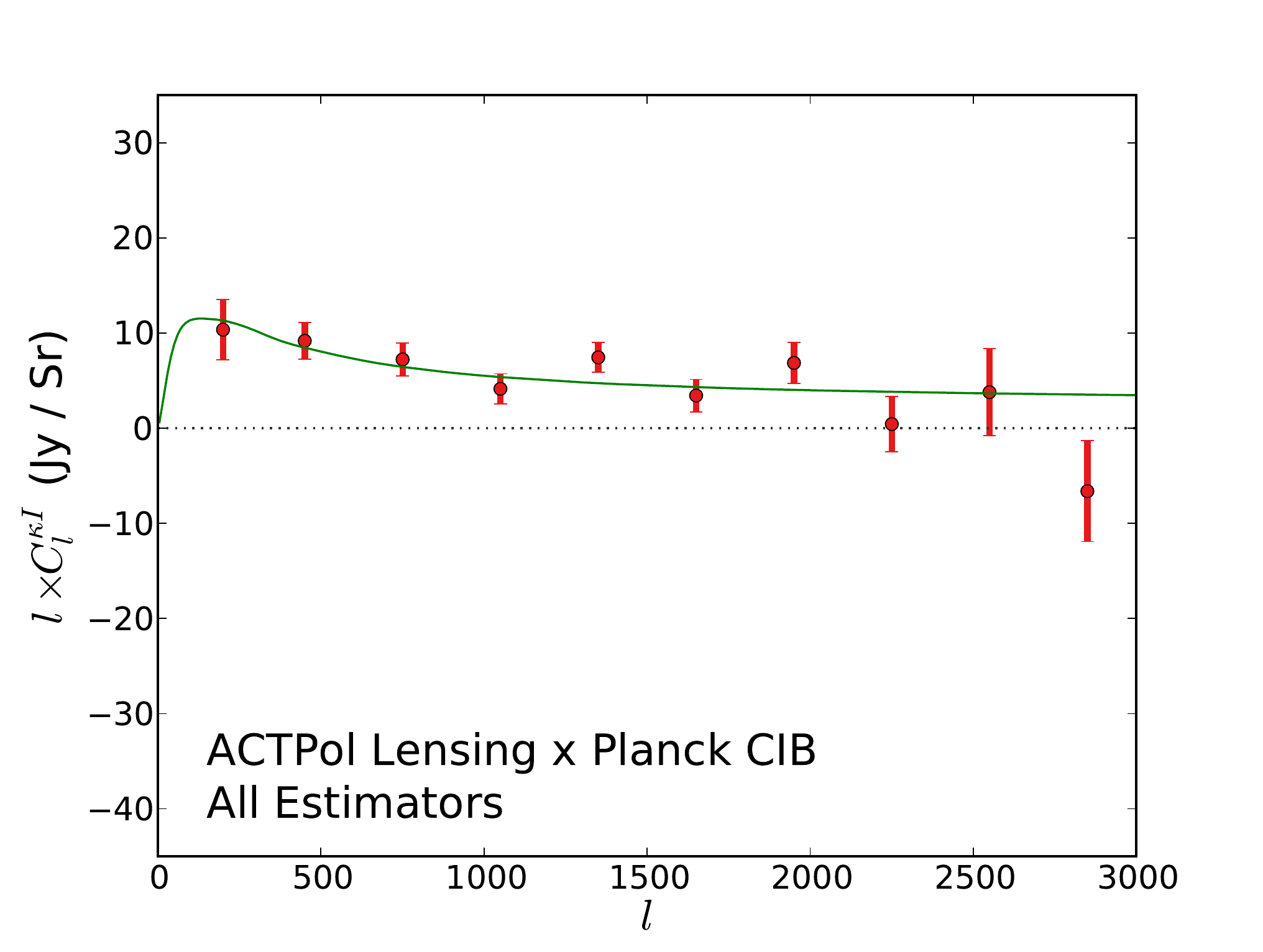}
\caption{Cross power spectrum of the reconstructed lensing convergence map from ACTPol data with a map of the CIB as measured by \planck\ at 545~GHz.  
 The power spectra from the combination of the TT, TE, EE, and EB estimators have been coadded for the D1, D5 and D6 sky regions.  The errors for each patch and estimator are determined from the cross power of 2048 simulated reconstructed lensing convergence maps with the appropriate \planck\ 545~GHz CIB map, and neighboring errors are less than 5\% correlated.  The detection significance of this lensing signal is $\AllpcSig \sigma$.  The green curve shown is not a fit to these data, but rather to the \citet{planck_ciblensing/2013} data.   We find a best-fit amplitude of $A = \AllpcBestfit$, with a chi-square statistic of $\AllpcChisquare$ for \dofsAllPatches\ degrees of freedom, and a probability to exceed the observed chi-square of \AllpcPte.  
  \vspace{3mm}}
\label{fig:plotAllPatches_Allpc}
\end{figure}

The lensing of the CMB can be detected by measuring the lensing-induced correlation between modes via an optimal quadratic estimator \citep{bernardeau/1997,seljak/1999, zaldarriaga/1999, hu/okamoto/2002} for the convergence field $\kappa$ given by  $- \frac{1}{2} \nabla \cdot {\bf d}$.  This estimator can be constructed from any pair of T, E, and B, denoted by $X$ and $Y$.  In the flat-sky limit, the estimator is given by 
\begin{equation}
\widehat{\kappa}_{XY}({\bf l}) = A_{XY}({\bf l})\int \frac{{\rm d}^2 {\bf l^\prime}}{(2\pi)^2}\, g^{XY}({\bf l^\prime},{\bf l - l^\prime})\, X({\bf l^\prime})\,Y({\bf l - l^\prime})
\end{equation}
where $g$ is a filtering function that optimizes the estimator, $A({\bf l}$ is a function that normalizes the estimate, and  ${\bf l^\prime }$ and ${\bf l}$ are  vectors  in two-dimensional Fourier space.  This estimator yields a convergence map derived from the $X$ and $Y$ fields.  In this analysis, we use only CMB modes above $l_{\rm{min}}=500$ and below $l_{\rm{max}}=\lmaxval$ for the T, E, and B fields.  The $l_{\rm{min}}$ limit is chosen to minimize bias from Galactic dust, and to match the choice made in N14.  The $l_{\rm{max}}$ limit is chosen to maximize signal-to-noise without introducing significant bias in the cross power spectrum from dusty galaxies correlated with the convergence field \citep{smith/etal/2007}.  We discuss the choice of $l_{\rm max}$ further in Section~\ref{sec:checks}.  We also remove a vertical strip in 2-dimensional CMB Fourier space with $| l_x | < 90$, and a horizontal strip with $| l_y | < 50$, as in N14.  

We  formulate a similar estimator for a field of ``curl'' deflections, $\widehat{\omega}({\bf L})$ \citep{hirata/seljak/2003, cooray/etal/2003}.  These curl deflections are expected to be more than two orders of magnitude smaller than the standard gradient-like deflections.  Cross-correlation of the reconstruction of this field with the CIB thus acts as an effective null test.

To validate the lensing pipeline, for each ACTPol patch we construct 2048 Monte Carlo simulations of the lensed CMB at 146~GHz. 
To lens simulated Gaussian T, Q, and U maps by the projected matter field we follow the procedure described in \citet{louis/etal/2014}.  
Random realizations of the noise are then added to the lensed CMB simulations. Gaussian noise realizations are generated from a template based on the 2-dimensional power spectrum of the noise, obtained by splitting the observations into four parts and differencing the resulting maps. To model the correct spatial inhomogeneity of the noise level, the noise realizations are then scaled according to the number of observations in each region of the map.

We use these simulations to construct a map of the window-induced mean field for each patch \citep{hanson/etal/2009, namikawa/etal/2009}, which we then subtract from both simulated and data-derived reconstructed convergence maps prior to any cross-correlation.  By cross-correlating the reconstructed convergence maps with the input convergence maps, we use the same simulations to obtain small $l$-dependent amplitude corrections due to the windowing, of $<5\%$, for which we correct.

\begin{figure}[t]
\includegraphics[width=1.05\columnwidth]{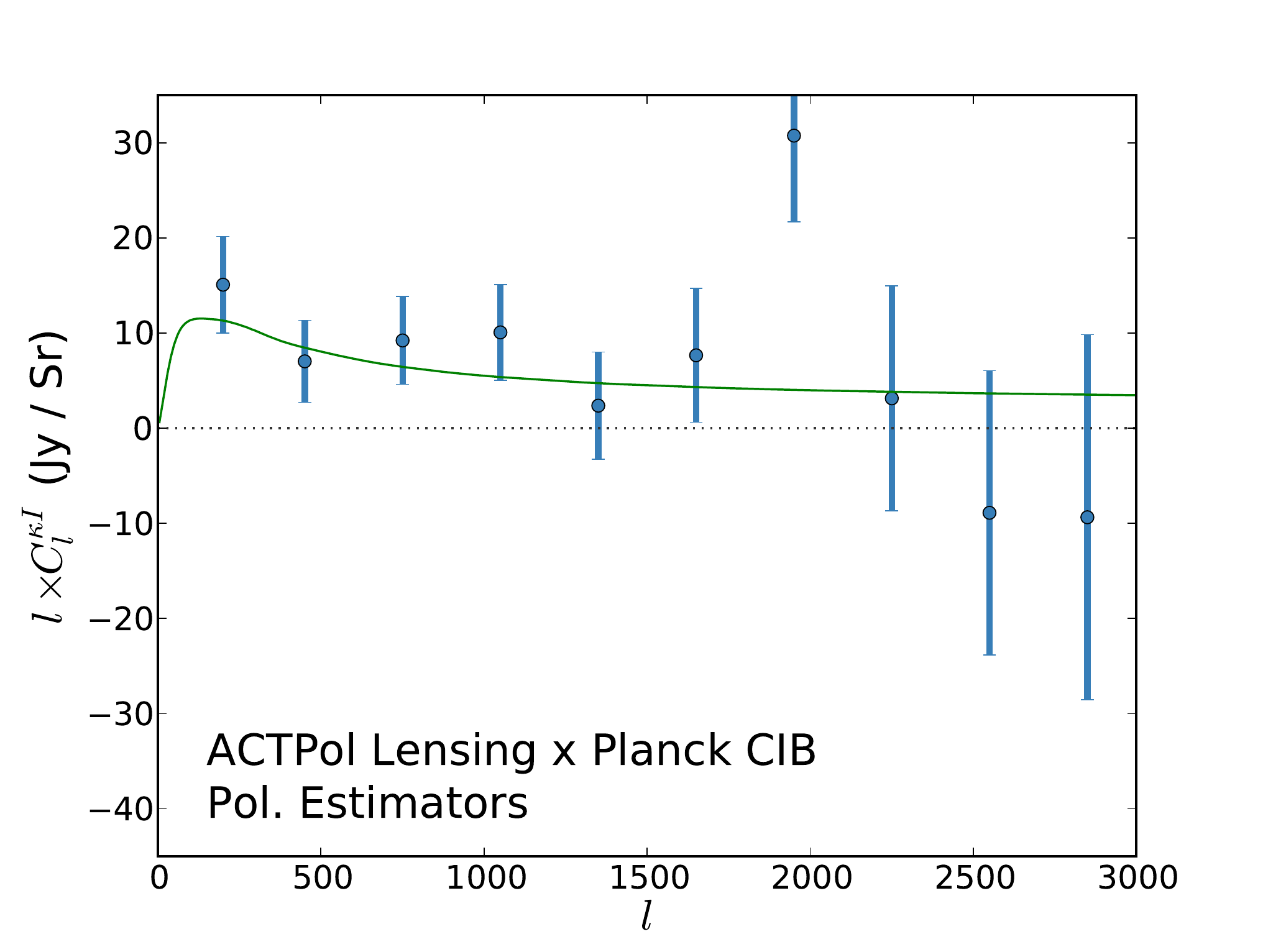}
\caption{Same as Figure \ref{fig:plotAllPatches_Allpc}, using only the EE and EB lensing estimators.  Measurements from the D1, D5, and D6 patches are  combined here as well.  The polarization lensing signal is detected at a significance of $\PolonlySig \sigma$.  Here we find a best-fit amplitude of $A = \PolonlyBestfit$, with a chi-square statistic of $\PolonlyChisquare$ for \dofsAllPatches\ degrees of freedom, and a probability to exceed the observed chi-square of \PolonlyPte. \vspace{3mm}}
\label{fig:plotAllPatches_Polonly}
\end{figure}

\section{Predicted Cross-Correlation}
\label{sec:prediction}
The CMB lensing convergence and the intensity of the CIB are expected to be correlated because both are tracers of the large-scale density fluctuations in the Universe.
We model the  cross-power as an integral over redshift \citep{song/etal/2003}:
\begin{equation}
C_l^{\kappa I} = \int dz\, {d\chi \over dz} {1\over\chi^2} W^\kappa(\chi) W^I(\chi) P\left(k = {
l \over \chi}, z\right).
\end{equation}
 Here $P(k,z)$ is the matter power spectrum including the effects of nonlinear growth  \citep{smith/etal/2003}.  We take the window function for the galaxies,  as a function of radial comoving distance $\chi$, to be
\begin{equation}
W^I(\chi) = {b \, {dI / d \chi} \over \int {d\chi\, dI / d \chi}}, 
\end{equation}
with a linear bias factor set to $b = 2.2$, which we assume to be independent of redshift, and with the CIB intensity distribution $I(\chi)$ at 500\,$\mu$m given by \citet{bethermin/etal/2013}.  The window function for CMB lensing is given by 
\begin{equation}
W^\kappa(\chi) = {3 \over 2} \Omega_m H_0^2 {\chi \over a(\chi)}  {\chicmb - \chi \over \chicmb 
}, 
\end{equation}
where $\Omega_m$ and $H_0$ are the matter density and the Hubble parameter today, $a(\chi)$ is the scale factor at comoving radial distance $\chi$, and $\chicmb \simeq 13\,$Gpc is the comoving distance to CMB recombination.

\begin{figure}[t]
\includegraphics[width=1.05\columnwidth]{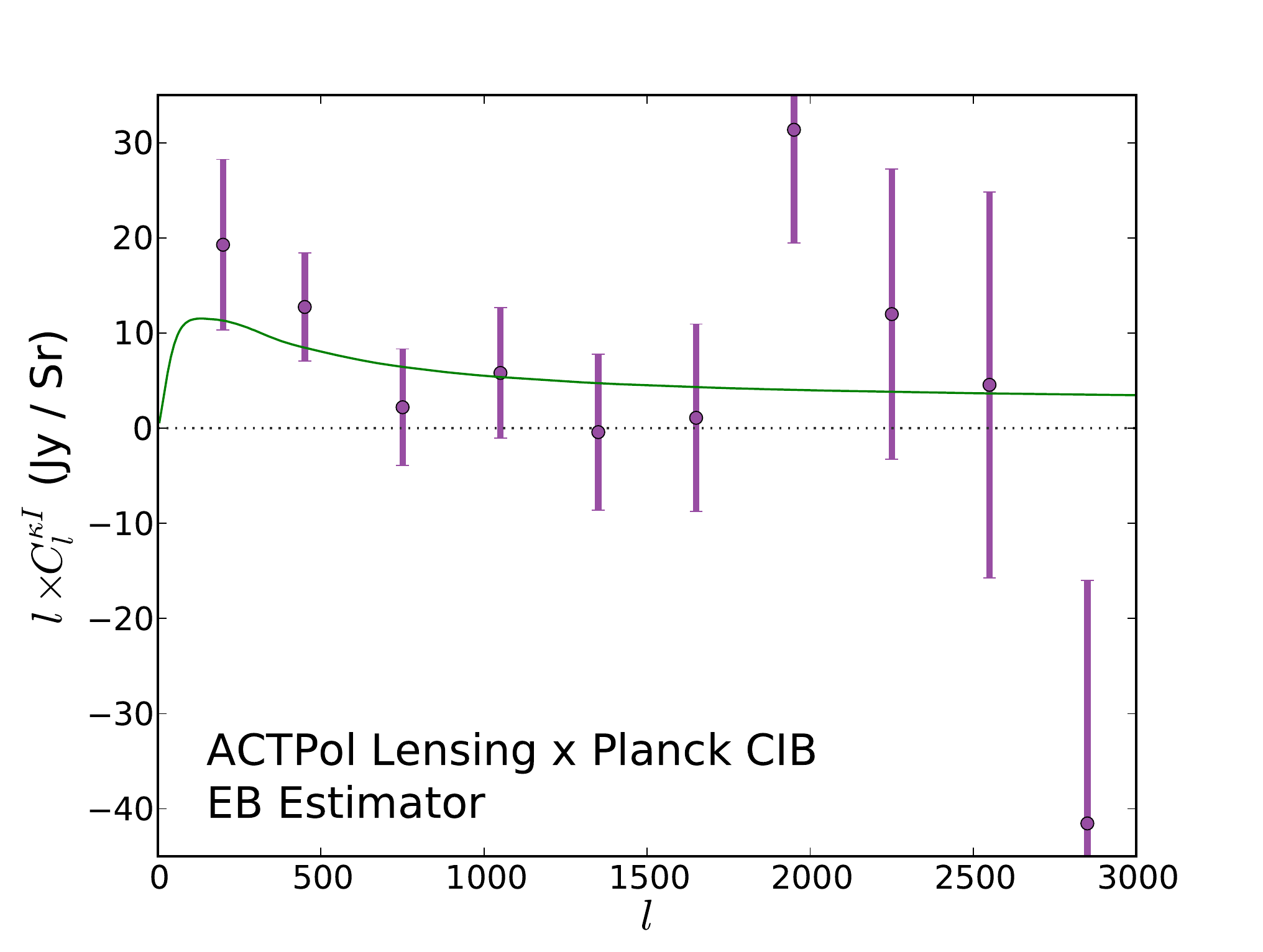}
\caption{Same as Figure \ref{fig:plotAllPatches_Polonly}, using only the EB lensing estimator. The B-mode lensing signal is detected at a significance of $\EBSig \sigma$.  Here we find a best-fit amplitude of $A = \EBBestfit$, with a chi-square statistic of $\EBChisquare$ for  \dofsAllPatches\ degrees of freedom and a probability to exceed the observed chi-square of \EBPte. \vspace{3mm}}
\label{fig:plotAllPatches_EB}
\end{figure}

\begin{figure}
\includegraphics[width=1.05\columnwidth]{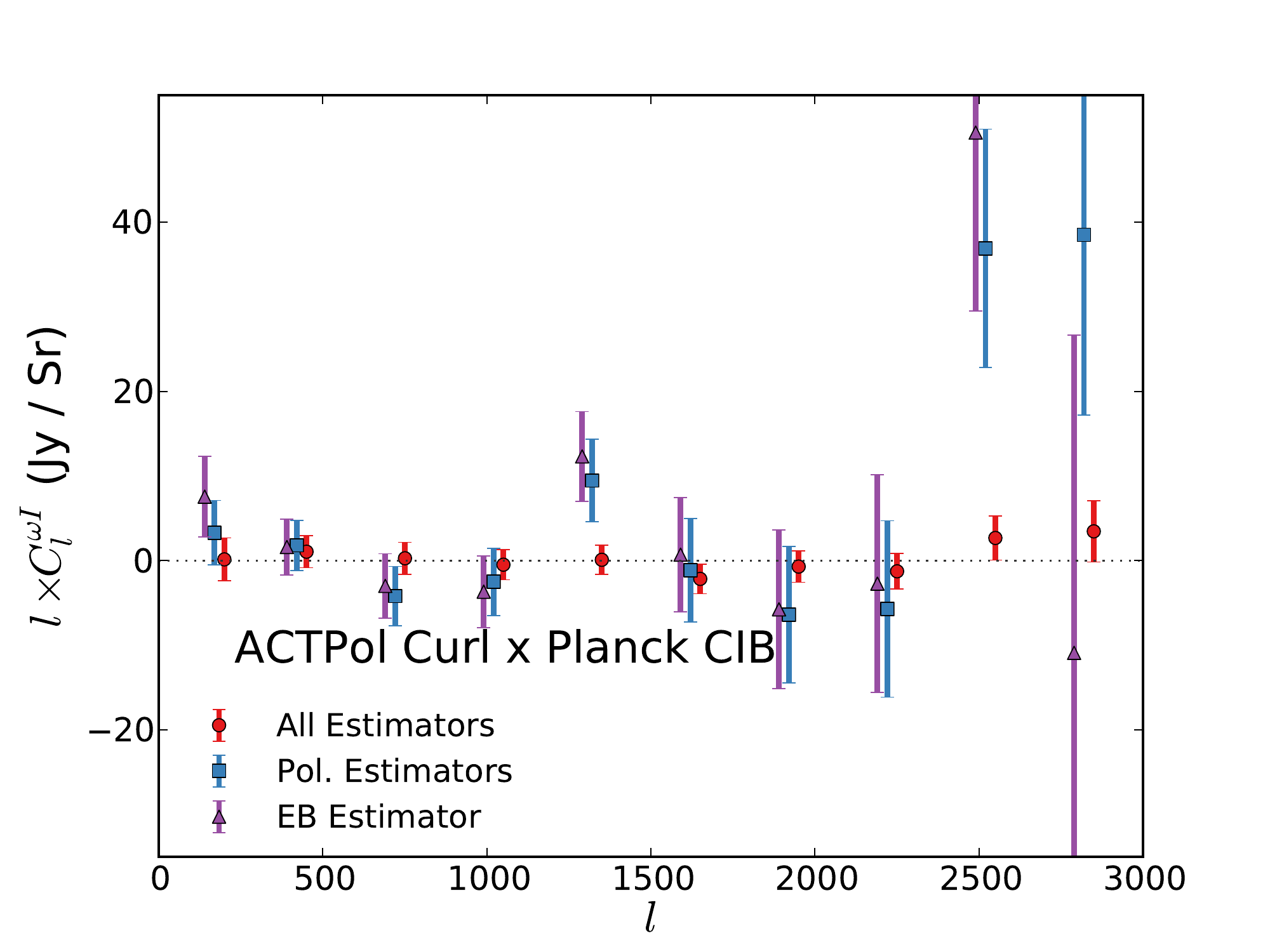}
\caption{Cross power spectra between estimates of the curl deflection field and the CIB. Estimator combinations are the same as in Figures \ref{fig:plotAllPatches_Allpc},\ref{fig:plotAllPatches_Polonly}, and \ref{fig:plotAllPatches_EB}.  All are consistent with null. \vspace{3mm}}
\label{fig:curlAltogether}
\end{figure}

This linear bias model
is an excellent match to the model that fits the recent CIB-lensing cross power measured by \planck\ \citep{planck_ciblensing/2013}. The \planck\  best-fit curve is based on a halo model which includes both the one and two-halo terms.  We note that our linear bias model, particularly the independence of the bias on redshift, is not meant to be a complete description of the CIB; it mainly serves as a simple model that matches existing data.

\section{Results}
\label{sec:results}

We obtain  estimated convergence maps for each patch using the TT, TE, EE, and EB lensing estimators, yielding 12 estimates of lensing.  We also obtain an equal number of estimates of the curl deflection field for the same data.  We additionally form combined estimates: a ``polarization-only'' combination, consisting of the EB and EE estimates combined; and an ``all'' combination, with all four estimates combined.  To obtain these combinations, we weight by the inverse variance of each estimator.

We then cross-correlate each of these reconstructed lensing and curl fields with the \planck\ CIB maps.  We obtain bandpower covariances from the 2048 Monte Carlo simulations discussed above, which we  correlate with the same \planck\ CIB maps.  We test for convergence of the covariance matrix by using half the simulations and checking that we obtain stable results.  The estimators are correlated \citep{hu/okamoto/2002}:  the simulated maps are generated from realizations of T, E, and B including the expected amount of TE cross-power, and  each simulated T, Q and U map is lensed by the same lensing field.  When combining estimators, we form the same linear combination with each of the 2048 simulations that we form with the data.  The inter-band covariance matrix which we obtain for each of these combinations thus includes all expected sources of correlation.

In Figure  \ref{fig:plotAllPatches_Allpc}, we show the result for the ``all'' combination of estimators, with error bars representing the on-diagonal part of the corresponding covariance matrix.  Neighboring bins are correlated by roughly $2$--$5\%$.  We fit to the model described in Section~\ref{sec:prediction} with a free overall amplitude, $A$.  
We find a best-fit amplitude of $A = \AllpcBestfit$,    corresponding to a detection  signal-to-noise ratio of $\sqrt{ (\chi^2_{{\rm null}} - \chi^2_{{\rm bf}}) }  = \AllpcSig$.  Here, $\chi^2_{{\rm null}}$ corresponds to the value of $\chi^2$ for $A = 0$, $\chi^2_{{\rm bf}}$ is the value for the best-fit point at which the $\chi^2$ is minimized, and we have summed the $\chi^2(A)$ curves over the three patches.
These results, together with those for each estimator and patch separately, are summarized in Table \ref{tab:vitalStats_kappa}.

We show the co-added cross power derived from the EE and EB estimators combined in Figure \ref{fig:plotAllPatches_Polonly}.  Here the lensing of the CMB polarization is detected at signal-to-noise ratio of $\PolonlySig$.  The cross power derived from the EB estimator alone is shown in Figure \ref{fig:plotAllPatches_EB}.  This yields a detection of B-mode polarization from lensing at a signal-to-noise ratio of $\EBSig$.  

We run the same pipeline for the curl reconstructions.  As shown in Table~\ref{tab:vitalStats_omega} and Figure \ref{fig:curlAltogether}, the curl estimates are consistent with null.  In Table~\ref{tab:vitalStats_omega}, we quote amplitudes relative to the usual scalar lensing curve.

\section{Systematic Error Tests}
\label{sec:checks}

The statistical uncertainties on our results for the cross-correlation amplitude are approximately $10\%$ for the TT estimator and $30\%$ for the EB estimator; we will show here that no known sources of systematic uncertainty are comparable in size.

\begin{figure*}[t]
\hspace{-0.7in}
\includegraphics[scale=.52]{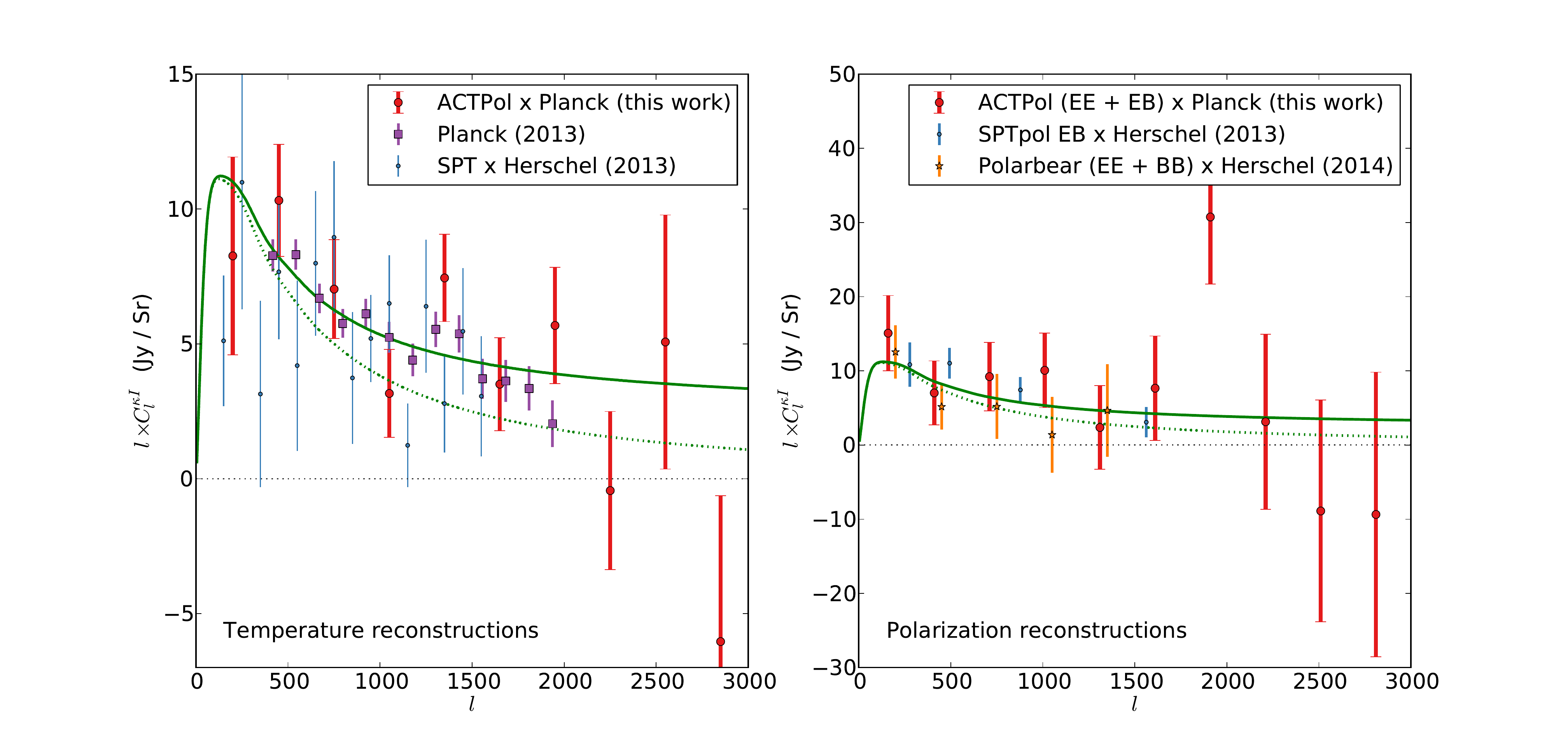}
\caption{Comparison with other surveys.  The left panel shows the temperature bandpowers from this work (red) together with those from the \planck\ lensing reconstruction cross-correlated with the \planck\ CIB maps at 545~GHz (purple), and the SPT lensing maps correlated with flux maps from \herschel at 500$\mu$m.  The right panel shows polarization results, with the results from this work (red), the \citet[][with EE and EB estimators combined]{pbear-herschel/2013} and \citet[][SPTpol, EB only]{hanson/etal/2013}.  All \herschel\ results have been color-corrected by a factor of \colorscaling\ to compare them to \planck\ CIB results which are at a different frequency.  
The green solid curve is as in Figs.1--3.  The dotted green curve shows the prediction using the linear matter power spectrum.}
\label{fig:comparison}
\end{figure*}

\begin{table}
\begin{center}
\begin{threeparttable}
\caption{Fits and $\chi^2$ values for the lensing convergence field}
\begin{tabular}{|l|c|c|r@{\,}l|c|}
\hline 
 & $S/N$ & $A$ & $\chi_{\rm bf}^2$ & ($ \nu $) & PTE\\ 
 \hline  
\hline 
TT, D1 &  2.4 & 0.90$  \pm 0.36  $ &   9.6 & (9) &  0.39\\   
TT, D5 &  3.3 & 0.82$^{+0.24}_{-0.20 }$ &  11.5 & (9) &  0.24\\   
TT, D6 &  7.2 & 1.06$  \pm 0.12  $ &  13.7 & (9) &  0.13\\   
 \hline 
 \hline 
TE, D1 &  1.5 & 2.10$  \pm 1.40  $ &   4.6 & (9) &  0.87\\   
TE, D5 &  0.2 & 0.22$^{+0.80}_{-0.84 }$ &  13.5 & (9) &  0.14\\   
TE, D6 &  2.0 & 0.90$  \pm 0.44  $ &   4.1 & (9) &  0.91\\   
 \hline 
 \hline 
EE, D1 &  1.3 & -2.14$^{+1.60}_{-1.56 }$ &   4.3 & (9) &  0.89\\   
EE, D5 &  0.5 & 0.46$  \pm 0.88  $ &   5.2 & (9) &  0.82\\   
EE, D6 &  4.0 & 1.74$^{+0.44}_{-0.40 }$ &  11.5 & (9) &  0.24\\   
 \hline 
 \hline 
EB, D1 &  0.1 & 0.26$^{+1.92}_{-1.88 }$ &   9.8 & (9) &  0.37\\   
EB, D5 &  1.3 & 1.26$^{+0.88}_{-0.92 }$ &   3.8 & (9) &  0.92\\   
EB, D6 &  2.9 & 1.38$^{+0.44}_{-0.48 }$ &  12.0 & (9) &  0.21\\   
 \hline 
 EB, all  &  3.2   & 1.30$  \pm 0.40  $ &  25.9 & (29) &  0.63\\ 
 \hline  
 \hline 
 Pol.\ estimators, all  &  4.5   & 1.26$^{+0.28}_{-0.24 }$ &  30.4 & (29) &  0.39\\ 
 \hline  
 \hline 
 All estimators, all  &  9.1   & 1.02$^{+0.12}_{-0.08 }$ &  37.2 & (29) &  0.14\\ 
 \hline  
\end{tabular} 
\vskip 2mm
\begin{tablenotes} \item  
\begin{center}
\begin{flushleft}
Fit results for the cross power between the lensing field from ACTPol maps and Planck maps at 545 GHz, for  each field and estimator.  The first column shows the signal to noise ratio calculated using the method described in the text.  The second shows the best-fit amplitude $A$, and  associated uncertainty, relative to the model which fits the \planck\ data.  The third column shows the values of $\chi^2$ at this best-fit point and the number of degrees of freedom.  The fourth shows the probability to exceed the given value of $\chi^2$.  The rows marked ``all'' are obtained by adding the $\chi^2$ functions and performing new fits for $A$.
\end{flushleft}
\end{center}
\end{tablenotes}
\label{tab:vitalStats_kappa}
\end{threeparttable}
\end{center}
\end{table}

Possible instrumental systematics include calibration and beam uncertainty. The largest known beam uncertainty is the overall (monopole) beam profile, for both the ACTPol and \planck\ surveys. For ACTPol, we estimate the effect of the beam and calibration errors together on the lensing reconstruction by computing effective error bands as functions of CMB multipoles, as described in N14.  This yields an effective error band across the CMB power spectrum of about $3\%$.  We then perform reconstruction on the temperature maps, scaling the maps in the Fourier domain by $+1$ and $-1 \sigma $.  This yields an offset of approximately $\pm 4\%$ in the best-fit cross-correlation amplitude for D6.  We thus assign a systematic uncertainty of $4\%$ to our final result.
The \planck\ fractional map-level beam error has amplitude 0.3\% to $l = 3000$ \citep{planck_beams/2013}, and is thus negligible for this analysis.

\begin{table}
\begin{center}
\begin{threeparttable}
\caption{Fits and $\chi^2$ values for the lensing curl field}
\begin{tabular}{|l|c|c|r@{\,}l|c|}
\hline 
 & $S/N$ & $A$ & $\chi_{\rm bf}^2$ & ($ \nu $) & PTE\\ 
 \hline  
\hline 
TT, D1 &  0.0 & 0.02$^{+0.36}_{-0.40 }$ &  17.2 & (9) &  0.05\\   
TT, D5 &  0.5 & -0.14$  \pm 0.24  $ &   6.7 & (9) &  0.67\\   
TT, D6 &  0.2 & 0.02$^{+0.16}_{-0.12 }$ &   8.8 & (9) &  0.46\\   
 \hline 
 \hline 
TE, D1 &  1.1 & -1.46$  \pm 1.32  $ &  11.9 & (9) &  0.22\\   
TE, D5 &  0.2 & 0.18$  \pm 0.80  $ &   5.7 & (9) &  0.77\\   
TE, D6 &  1.1 & -0.50$^{+0.40}_{-0.44 }$ &   5.6 & (9) &  0.78\\   
 \hline 
 \hline 
EE, D1 &  0.1 & 0.26$^{+1.96}_{-1.92 }$ &   3.9 & (9) &  0.92\\   
EE, D5 &  0.5 & 0.54$^{+1.04}_{-1.00 }$ &   7.2 & (9) &  0.62\\   
EE, D6 &  1.6 & -0.78$^{+0.48}_{-0.52 }$ &  11.5 & (9) &  0.24\\   
 \hline 
 \hline 
EB, D1 &  1.9 & 2.26$  \pm 1.16  $ &  10.9 & (9) &  0.28\\   
EB, D5 &  1.1 & 0.66$  \pm 0.56  $ &   5.5 & (9) &  0.79\\   
EB, D6 &  0.1 & 0.02$^{+0.28}_{-0.24 }$ &  17.3 & (9) &  0.04\\   
 \hline 
 EB, all  &  1.0   & 0.22$^{+0.24}_{-0.20 }$ &  37.8 & (29) &  0.13\\ 
 \hline  
 \hline 
 Pol.\ estimators, all  &  0.5   & 0.10$  \pm 0.20  $ &  32.7 & (29) &  0.29\\ 
 \hline  
 \hline 
 All estimators, all  &  0.1   & 0.02$^{+0.08}_{-0.12 }$ &  36.3 & (29) &  0.16\\ 
 \hline  
\end{tabular} 
\vskip 2mm
\begin{tablenotes} \item  
\begin{center}
\begin{flushleft}
Null check for the cross power between the curl lensing field obtained from ACTPol maps and Planck maps at 545 GHz, for  each field and estimator.  Columns are as in Table \ref{tab:vitalStats_kappa}, where quantities are quoted relative to the same (scalar) model.
\end{flushleft}
\end{center}
\end{tablenotes}
\label{tab:vitalStats_omega}
\end{threeparttable}
\end{center}
\end{table}

Leakage from temperature to polarization could potentially affect our polarized lensing signal. To estimate the effect, we simulate leakage by adding $1\%$ of the temperature map to Q and U, as well as $ 10\% $ of Q to U (and vice versa). We then propagate these new maps through the lensing estimator and cross-correlate them with simulated, correlated CIB maps. We find only a small contamination, of order 2\% of the signal amplitude for all the polarized lensing channels. This is consistent with previous results \citep{pbear-herschel/2013}, where the effect of leakage was shown analytically to be zero in the EB channel, and the effects of leakage in other channels were found to be subdominant.

Errors in the polarization angles might have an impact on the measurement. Performing a $1$ degree rotation of the polarization angle in simulated maps and propagating the maps through our lensing reconstruction and cross-correlation pipeline, we find a change of only 2\% in the cross-correlation signal measured. Given that our bound on the angle error is $0.5^\circ$ from measurements of null cross-power (N14), we conclude that this source of error is negligible for our purposes.

We now turn to estimates of astrophysical systematic errors. An important source of possible systematic contamination is flux from dusty sources in the CMB maps, which propagates through the lensing estimator and forms a non-zero bispectrum when the lensing estimator is correlated with the \planck\ maps.  This signal is proportional to the $\EV{I_{\rm 150} ({\bf l}_1) I_{\rm 150} ({\bf l}_2)I_{\rm 545}({\bf l}_3)}$ three-point function for three multipole vectors ${\bf l}_1$, ${\bf l}_2$, and ${\bf l}_3$.  For spatially uncorrelated sources, the associated bispectrum is constant in $l$ and is small ($<1\%$) with the point-source flux threshold we have applied \citep{vanengelen/etal/2014, planck_ciblensing/2013}. 

The clustered CIB bispectrum can also give spurious lensing signals \citep{planck_ciblensing/2013}.  The first detections of clustering in the CIB bispectrum have recently been made \citep{crawford/etal/2014, planck_cib/2013}, and simulations give varying levels of clustered bispectra \citep[e.g.,][]{vanengelen/etal/2014}.  A bias term arising from a three-point function of the form $\langle \kappa({\bf l}_1) I({\bf l}_2) I({\bf l}_3) \rangle$ was shown by \citet{vanengelen/etal/2014} and \citet{osborne/etal/2013} to affect cross-correlation analyses at the level of a few percent.  Preliminary analysis of the full CIB bispectrum using the simulations analyzed by \citet{vanengelen/etal/2014}, including those of \citet{sehgal/etal/2010},  appears to yield large biases (tens of percents), depending on the bispectrum level, the flux threshold,  the masking method, and the maximal CMB multipole $l_{\rm max}$.  Given the dependence of this signal on these variables, and the uncertainties of CIB modelling, we take an empirical approach in determining this source of bias for our measurement. 

Given that extragalactic foreground biases arise from structure in the CMB maps at high $l$, we rerun our entire analysis with $l_{max} = 2000$.  We find that the best-fit cross-correlation amplitude for all estimators combined shifts to $A = \AllpcBestfitTwothou$.  Conversely, setting $\lmax = 4000$ leads to a best-fit amplitude of $A = \AllpcBestfitFourthou$.  While we note the trend that $A$ shifts down as $\lmax$ increases, given our current errors we cannot claim direct evidence of a foreground bias.

To further investigate the potential bias from CIB, including clustering,  we construct an estimate of the CIB at 146 GHz by scaling down the beam-deconvolved 545 GHz \planck\ map assuming a spectral index of $\alpha = 2.75$ \citep{gispert/etal/2000}.  We then calculate the expected bias  by first performing lens reconstruction on this map using the TT estimator, and then cross-correlating the result with the unscaled 545 GHz map from our nominal analysis.  This approach is valid because other cross-terms in the three-point function are proportional to a single factor of the  primordial (unlensed) CMB field and hence are zero on average.   For D6, the cleanest patch, the shift is $<1\%$; for D5 it is $8\%$; and for D1, the dustiest patch, the shift is 40\%.  
Since a CIB bias should be uniform  across fields, whereas dust is much less isotropic, we attribute this to the higher dust levels in D1 compared to the other fields, rather than to the CIB bispectrum.  A similar argument was presented by the \citet{planck_ciblensing/2013}.  We conclude that the clustered CIB bispectrum should be negligible for our analysis.

To estimate the impact of Galactic dust, we use two templates: the high-frequency, beam-deconvolved \planck~ maps, as well as the maps of  \citet{finkbeiner/davis/schlegel/1999}.  We again scale the maps at 545 GHz to 146 GHz, but here we assume a dust spectral index of $\alpha = 3.18$ \citep{planck_thermaldust/2013}, as we have argued that the CIB portion of these maps does not cause a bias.  We assume a polarization fraction of 20\%, commensurate with some of the most polarized regions of the mm-wave sky at high Galactic latitude.  We thus set $Q = 0.2 T_{\rm 545}$ and $U = 0$, where $T_{\rm 545}$ is the rescaled \planck~map.  The resulting shifts for the TT estimator are  12\% for D1,   4\% for D5, and $<1\%$ for D6.  The total shift in the TT estimators, using our weighting, is 4\%.   The polarized estimators show small biases, of 8\% for the EB estimator (compared with a 30\% statistical error bar), and $<1\%$ for the TE and EE estimators.  Using the \citet{finkbeiner/davis/schlegel/1999} maps, and the same treatment for polarization, we find a $\sim 3\%$ bias for the EB estimator in D1, and $\lesssim 1\%$ for the other fields and estimators, including TT.  This difference may be due to the lower angular resolution of the \citeauthor{finkbeiner/davis/schlegel/1999} maps.  Given the difference between these two results, we assign a  systematic uncertainty on our measurement due to dust contamination of 4\%.  We add this in quadrature to the beam and calibration uncertainty, yielding a 6\% systematic uncertainty.

\begin{figure}[t]
\includegraphics[width=1.05\columnwidth]{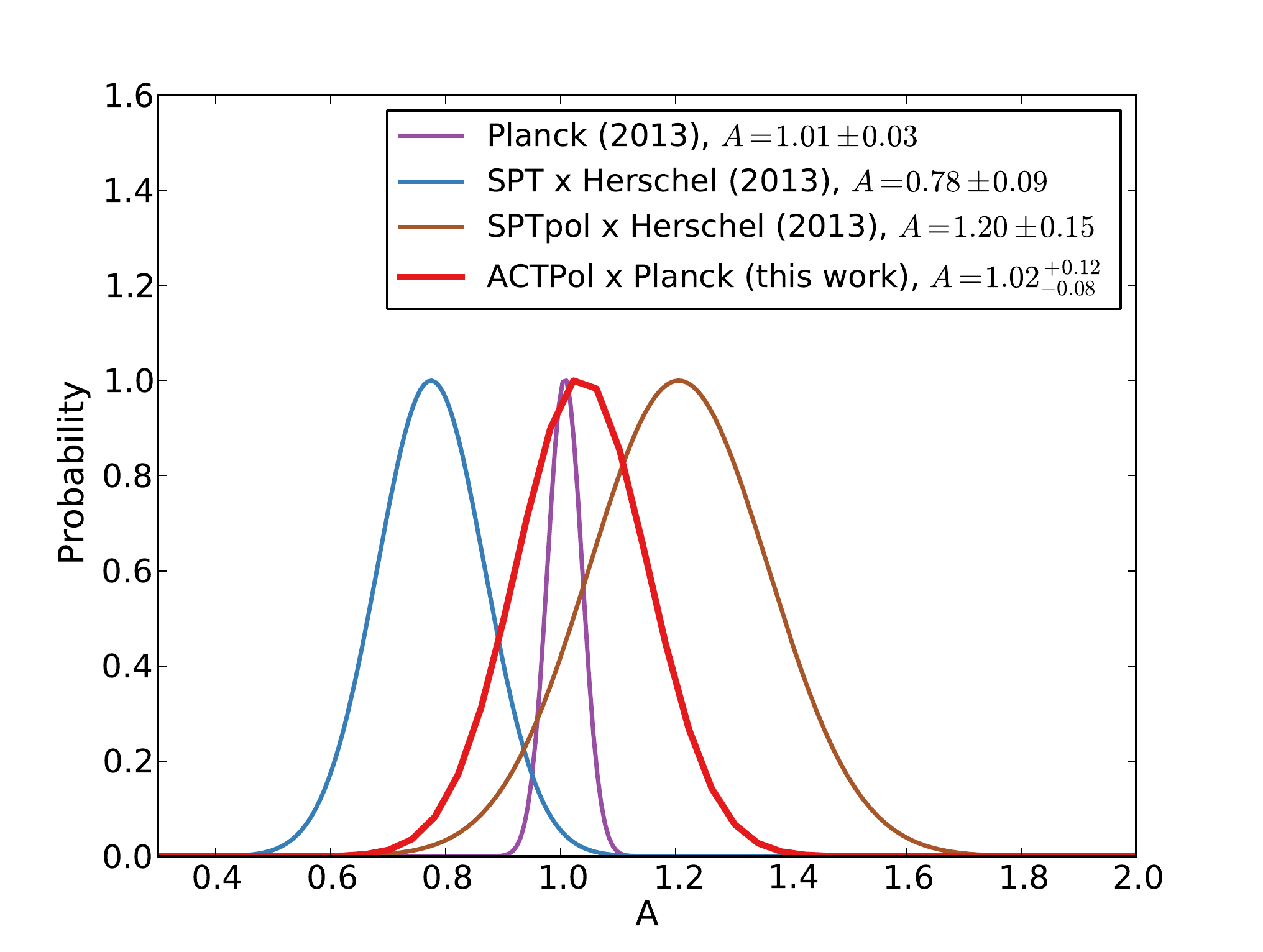}
\caption{Amplitude comparison of cross-correlation of CMB lensing with CIB emission at 545 GHz (\planck) and 500\,$\mu$m (600 GHz; \herschel) from different experiments.  Shown are fits to bandpowers from Figure~\ref{fig:comparison}. We only treat statistical uncertainties in this plot.  All \herschel~results have been scaled downwards by a factor of \colorscaling.}
\label{fig:amplitudeComparison}
\end{figure}

\section{Consistency with other surveys}
\label{sec:consistency}
The cross power spectrum between the CIB and CMB lensing has now been measured by several groups.  In Figure~\ref{fig:comparison}  we include results from the \planck\ lensing reconstruction cross-correlated with the \planck\  maps at 545~GHz  \citep{planck_ciblensing/2013}, an SPT temperature-based lensing map cross-correlated with a \herschel-SPIRE (\herschel~hereafter)  map at 500~$\mu$m \citep{holder/etal/2013}, an SPTpol polarization-based lensing map cross-correlated with \herschel~ at 500~$\mu$m \citep{hanson/etal/2013}, and the \textsc{Polarbear} polarization-based lensing maps cross-correlated with \herschel\ at 500~$\mu$m \citep{pbear-herschel/2013}.  

To estimate the consistency of the data, we fit the bandpowers of the four datasets that have a detection significance greater than $5\sigma$ to the model curve considered in this paper, which is an excellent fit to the \planck\ bandpowers.  We scale all \herschel-SPIRE $500\,\mu$m (600~GHz) results down by a color-correction factor of \colorscaling\ \citep{planck_cib/2013, gispert/etal/2000} for comparison with \planck\ results at 545~GHz.  We assume a simple $\chi^2$ likelihood, and do not include any correlations between bands.  We do not include calibration errors for \herschel\ $500\,\mu$m or \planck\ 545~GHz data as the relative calibration between these maps was shown to be within $3\%$  \citep{planck_calibration/2013}.    The results are shown in Figure~\ref{fig:amplitudeComparison}.  Using this color-correction factor (as opposed to the factor 1.22 used by \citealt{hanson/etal/2013}),  the surveys are generally consistent; the SPT result is in mild tension with the Planck result at the $\sim 2\sigma$ level.  The best-fit amplitude found in this work is broadly consistent with all the surveys.

\section{Effective ACTPol B-mode level}
The measurements of $C_l^{\kappa I}$ shown in Figs 1 through 4 represent one way of showing the lens-induced three-point function between measured CMB E modes, CMB B modes, and the CIB intensity field.  Since the fluctuation amplitudes of both the CMB E modes and the CIB intensity field are already well measured, specifically to amplitudes of 0.5\% \citep{planck_cib/2013} and {2.6}\% (N14) respectively, a detection of the cross-correlation using the EB estimator can be cast as a detection of lens-induced B modes.  Indeed, \citet{hanson/etal/2013} interpreted their measurement of this bispectrum as the first evidence for B modes induced by lensing \citep{smith/etal/2007}.   Our EB lensing estimator shows B-mode lensing at a significance of  $\EBSig  \sigma$.  This result is shown, together with other recent measures of the  B-mode power spectrum, in Figure \ref{fig:clbbPlot}.  We have treated the  ACTPol measurement as a single effective bandpower in the B-mode power spectrum with amplitude relative to fiducial of $A = \EBBestfit$.  We place this single effective bandpower at  $l = 1000$, near the center of the distribution for lensing information with the EB lensing estimator \citep{pearson/etal/2014}.

\begin{figure}[t]
\includegraphics[width=1.05\columnwidth]{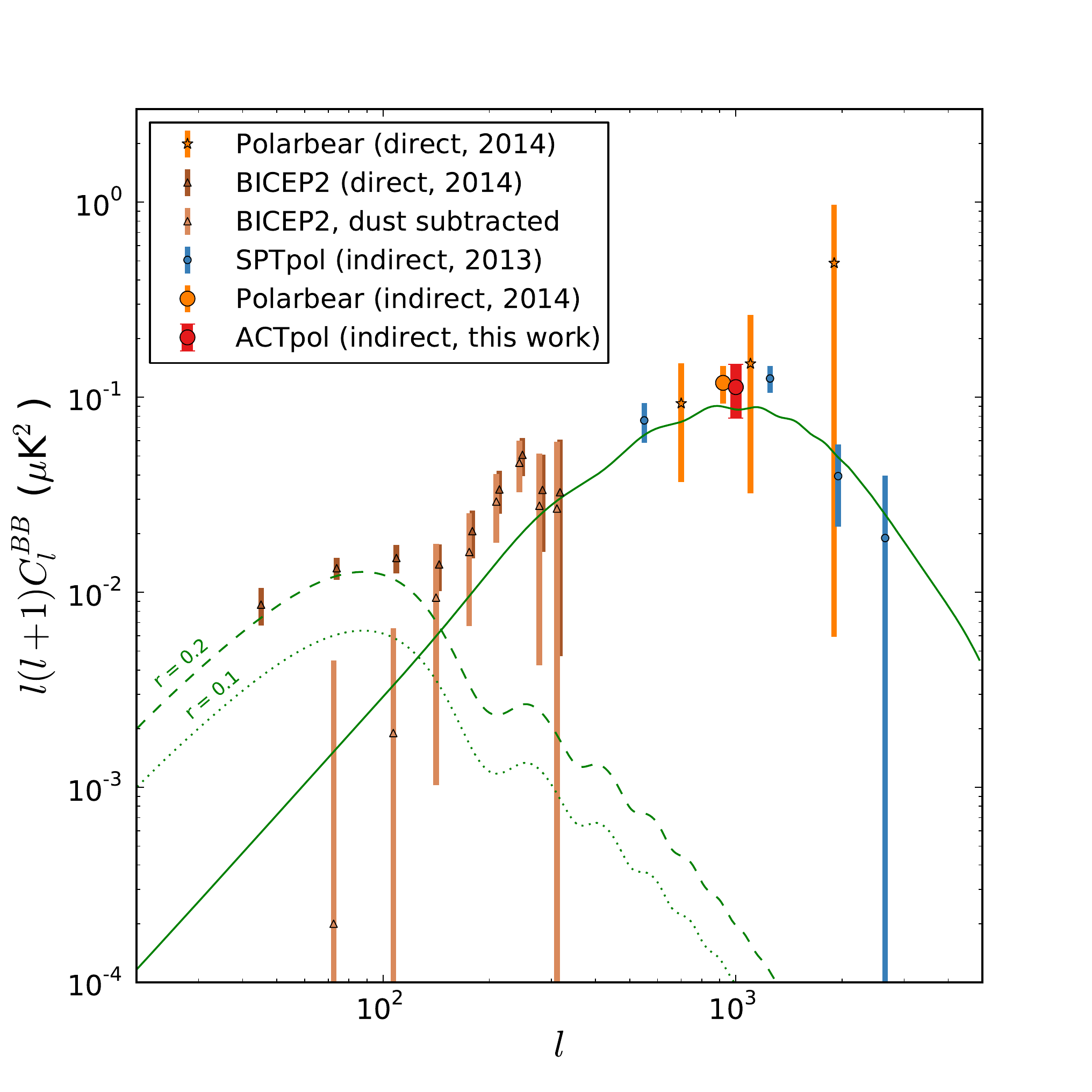}
\caption{Recent measures of B-mode power at 150 GHz.  The solid green curve is the expectation for lens-induced B modes, while the dashed and dotted lines show the expectations for primordial gravitational waves for two reference amplitudes.  Shown are direct measures of the B-mode power spectrum $C_l^{BB}$, including those by the \citet[][orange stars]{pbear-eebb/2014} and the \citet[][dark brown diamonds]{bicep2a/2014}.   We estimate the lensing B-modes from the BICEP2 power spectrum measurement by subtracting the central values of dust contamination at 150 GHz given in Figure 9 of \citet{planck_dustbicep/2014}  (light brown diamonds), which are roughly comparable   to the tensor curves shown.  Also shown are indirect B-mode measures obtained from the amplitude of B modes arising from the CMB lensing-CIB correlation, including results from SPTpol \citep[][blue squares]{hanson/etal/2013}  and this work (red circle).  The orange circle is a similar result using lensing autospectra from B-mode estimates, from \citet{pbear-eeeb/2013}. The ACTpol point reflects the measured amplitude $A = \EBBestfit$ relative to the fiducial model.  \vspace{3mm}}
\label{fig:clbbPlot}
\end{figure}

\section{Summary}
\label{sec:discussion}
We have presented the first large-scale lensing results from ACTPol, a polarization-sensitive camera on the ACT telescope, using the cross-correlation between the lensing field and another tracer of large-scale structure, the unresolved galaxies comprising the Cosmic Infrared Background.  Using the first 600 hours of data from  the ACTPol survey, we have demonstrated lensing of the CMB polarization at $\PolonlySig \sigma$, and a  $\AllpcSig \sigma$ detection including the temperature data. Lensing cross-correlations are thus emerging as strong probes of the manner in which galaxies trace mass in the Universe.  The CIB is also promising as a  proxy for the lensing field for the purpose of  removing the lens-induced B modes in searches for primordial gravitational waves \citep{simard/etal/2014}.   

The ACTPol survey is now in its second season and is observing with an upgraded receiver.   As more CMB polarization data are obtained in the near future, lensing of the CMB polarization  will become a powerful probe of precision cosmology.

\acknowledgments
We thank Duncan Hanson for discussion and clarification regarding Section~\ref{sec:consistency}.  We additionally thank Olivier Dor\'{e}, Gil Holder, Guilaine Lagache, and Marco Viero for useful correspondence.  

This work was supported by the U.S. National Science Foundation through awards
AST-0408698 and AST-0965625 for the ACT project, as well as awards PHY-0855887
and PHY-1214379. Funding was also provided by Princeton University, the
University of Pennsylvania, Cornell University, the University of Michigan, and a Canada Foundation for Innovation (CFI) award
to UBC. AK is supported by NSF grant AST-13122380.  We gratefully acknowledge support from the Misrahi and Wilkinson research funds.  We acknowledge Oxford ERC grant 259505.   The development of
 detectors and lenses was supported by NASA grants NNX13AE56G and
 NNX14AB58G.  We also acknowledge support from CONICYT grants QUIMAL-120001 and FONDECYT-1141113. ACT operates in the Parque Astron\'omico Atacama in northern Chile
under the auspices of the Comisi\'on Nacional de Investigaci\'on Cient\'ifica y
Tecnol\'ogica de Chile (CONICYT). Computations were performed on the GPC
supercomputer at the SciNet HPC Consortium. SciNet is funded by the CFI under
the auspices of Compute Canada, the Government of Ontario, the Ontario Research
Fund -- Research Excellence; and the University of Toronto. 

\bibliographystyle{act}
\bibliography{lenscib_refs.bib,apj-jour}

\end{document}